\documentclass{elsart}
\usepackage{epsfig}
\usepackage{amsmath,amsfonts}
\newcommand{\beq}{\begin{equation}}
\newcommand{\eeq}{\end{equation}}
\newcommand{\bear}{\begin{eqnarray}}
\newcommand{\ear}{\end{eqnarray}}
\newcommand{\earn}{\nonumber \end{eqnarray}}
\newcommand{\dst}{\displaystyle}
\newcommand{\nn}{\nonumber \\}

\newcommand{\gsim}{\mathop{\lefteqn{\raise.9pt\hbox{$>$}}
\raise-3.7pt\hbox{$\sim$}}}
\newcommand{\llsim}{\mathop{\lefteqn{\raise.9pt\hbox{$<$}} \raise-3.7pt\hbox{$\sim$}}}
\arraycolsep=\mathsurround

\begin{document}

\title{Self-force on a scalar point charge in the long throat}

\author{Arkady A. Popov}
\address{
Department of Mathematics, Tatar State University of Humanities and Education, Tatarstan 2, Kazan 420021, Russia\\
Department of Physics, Kazan State University, Kremlevskaya 18, Kazan, 420008, Russia}
\ead{apopov@ksu.ru}

\begin{abstract}
An analytic method is presented which allows for the computation
of the self-force for a static particle with a scalar charge in
the region of an ultrastatic spacetime which one can call the long
throat. The method is based on the approximate WKB solution of a
radial mode equation for a scalar field. This field is assumed to
be massless, with a coupling $\xi$ to the scalar curvature is
satisfying the condition $\xi>1/8$.
\end{abstract}


\maketitle


\section{Introduction}


The study of a self-force has a long history. The original
investigations focused on the self-acceleration of an
electrically-charged point particle in flat spacetime
\cite{Dirac:1938}. Later DeWitt, Brehme, and Hobbs
\cite{DeW-Bre:1960} studied the influence of the self-force on a
charge in a curved spacetime. In contrast to the case of a flat
spacetime this force can be nonzero even for a static charge in a
curved background. A number of static configurations has been
analyzed, including the self-action in the spacetimes of a
Schwarzschild black hole \cite{SmithWill:1980,Lohiya:1982}, of a
Kerr black hole \cite{LeauteLinet:1982}, of a Kerr-Newman black
hole \cite{Lohiya:1982} and  in a spherically symmetric
Brans-Dicke field \cite{LinTeys:1979}. The analytic approximation
of self-force has been obtained for a scalar charge at rest in an
axisymmetric spacetime \cite{BurkoLiu:2001}. The self-force can be
nonzero for a static particle in flat spacetimes of the
topological defects \cite{Linet:1986}. In curved spacetimes with
nontrivial topological structure the investigations of this type
have the additional interesting features
\cite{KhusBakh:2007,BezerraKhus:2009}.

The effect of self-action is associated with nonlocal structure of
the massless field, the source of which is the charged particle.
For example, the self-force on a scalar charge is \cite{Quinn:2000}
\bear
f_\mu &=& q^2 \left[ \frac{1}{3} \left( \dot{a}_\mu - a^2 u_\mu \right)
+ \frac{1}{6} \left( R_\mu^{ \nu} u_{\nu}
+ R_{\nu \gamma} u^\nu u^\gamma u_\mu \right)
\right. \nn && \left.
+ \frac{1}{12} \left( 6 \xi-1 \right) R \, u_\mu
+ \lim _{\epsilon \rightarrow 0} \int_{-\infty}^{\tau
- \epsilon} \nabla_\mu G_{ret} \left( x,x' \right) d\tau '\right]
\ear
where $u_\mu$ is the 4-velocity of the particle, $a_\mu$ is the 4-acceleration,
$\dot{a}_\mu = {\partial{a_\mu}}/{\partial{\tau}}$ is the derivative of
the 4-acceleration with respect to proper time $\tau$ of a charged particle,
$G_{ret}(x,x')$ is the retarded scalar Green's function and $\xi$ is
the coupling to the background scalar curvature.

Are there the situations in which the effect of self-action is determined by
the local geometry of the curved spacetime?
As it is demonstrated below such a situation for the static scalar charge
takes place, for example, in the throat of the wormhole if the length of this
throat is much more than the radius of throat. As the examples of such wormholes
one can consider the spacetimes with metric
\beq \label{met1}
ds^2= -d t^2+d\rho^2
+\left( r_0+ \rho \tanh \frac{\rho}{\rho_0} \right)^2
\left(d\theta^2+\sin^2\theta\, d\varphi^2 \right)
\eeq
or
\beq \label{met2}
ds^2= -d t^2+d\rho^2
+\left( r_0+ \rho \coth \frac{\rho}{\rho_0}-\rho_0 \right)^2
\left(d\theta^2+\sin^2\theta\, d\varphi^2 \right),
\eeq
where $r_0, \ \rho_0$ are the constants ($r_0$ is a radius of the
throat, $\rho_0$ is the parameter which describes the length of the throat)
and
\beq  \label{c}
\frac{r_0}{\rho_0} \ll 1.
\eeq
The effect of self-action in the region $\rho \llsim \rho_0$ does
not depend on the geometry of a spacetime outside of this region
and we shall call this region the long throat
(the accurate determination of the long throat is given below).

The organization of this Letter is as follows.
In the following section we develop the general approach to
a procedure of the self-force calculation.
In section III, we develop an approximation for the self-force
acting on a scalar charge at fixed position using the WKB approximation for
the radial modes of the scalar field.
In section IV, we evaluate the explicit expressions
for the self-force on the two specific gravitational backgrounds.
Finally, in section V we present the concluding remarks.

Throughout this Leter we use units $c=G=1$.

\section{General approach}

Let us consider a massless scalar field $\phi$ with scalar source $j$. The
corresponding action is given by
\beq
 S = -\frac{1}{8\pi} \int \left(\phi_{,\mu}\phi^{,\mu} + \xi R \phi^2 \right)
\sqrt{-g} \, d^4x + \int j\phi  \sqrt{-g} \, d^4x,
\eeq
where $\xi$ is a coupling of the scalar field to the scalar
curvature $R$ and $g$ is the determinant of the metric
$g_{\mu \nu}$. The corresponding field equation has a form
\beq
\left(\frac{}{}\square_x -\xi R(x)\right)\phi(x; \tilde x)
= -4\pi j (x; \tilde x),
\eeq
where
\beq \label{j}
j(x; \tilde x) =  q \int \delta^{(4)}(x^\mu,{\tilde x}^\mu (\tau))
\frac{d\tau}{\sqrt{-g}},
\eeq
is the scalar current,
$q$ is the scalar charge and $\tau$ is its proper time.
The world line of the charge is given by ${\tilde x}^\mu (\tau)$.
We shall consider only the case in which the charge is at rest in
an ultrastatic spacetime.
This means that one can rewrite the field equation in the following way
\beq \label{fieldeq}
\left(\frac{}{}\triangle_x -\xi R(x^\alpha)\right)
\phi(x^\alpha; \tilde x^\alpha)
= -  \frac{4\pi q}{u^t\sqrt{-g}} \,
\delta^{(3)}(x^\alpha,\tilde x^\alpha),
\eeq
where $t$ is the time coordinate, $u^t=dt/d\tau$ and $\alpha=1, 2, 3$.

The procedure of the self-force evaluation requires the renormalization
of a scalar potential $\phi(x; \tilde x)$ which is diverged in
the limit $x \rightarrow \tilde x$ (see, for example, \cite{KhusBakh:2007,BezerraKhus:2009}).
This renormalization is achieved by subtracting from $\phi(x; \tilde x)$
the DeWitt-Schwinger counterterm $\phi_{\mbox{\tiny \sl DS}}(x; \tilde x)$ and then
letting $x \rightarrow \tilde x$:
\beq
\phi_{\mbox{\tiny \sl ren}}(x)=\lim_{ \tilde x \rightarrow x}
\left[ \phi(x; \tilde x)- \phi_{\mbox{\tiny \sl DS}}(x; \tilde x)\right].
\eeq
The expression for $\phi_{\mbox{\tiny \sl DS}}(x; \tilde x)$
is evaluated in \cite{Chr78} and in $3D$ case has a form (see, also, \cite{BezerraKhus:2009})
\beq
\phi_{\mbox{\tiny \sl DS}}(x; \tilde x)=q \frac{\triangle^{1/2}}{\sqrt{\, 2\sigma}}
\eeq
where $\sigma$ is one-half the square of the distance between the points $x$
and $\tilde x$ along the shortest geodesic connecting them and
$\triangle$ is DeWitt-Morrett determinant.

Finally the self-force acting on a static charge is
\beq
f_\alpha(x)=-\frac{q}{2}\nabla_\alpha \phi_{\mbox{\tiny \sl ren}}(x).
\eeq

\section{WKB approximation for the self-force }

The metric of an ultrastatic spherically symmetric spacetime under
consideration is
\beq\label{metric}
ds^2= -d t^2+d\rho^2+r^2(\rho)\left(d\theta^2+\sin^2\theta\, d\varphi^2 \right).
\eeq
In this spacetime for the static charge $u^t=1$ the solution of (\ref{fieldeq})
can be expanded in terms of Legendre polynomials $P_l$ with
the result that
\beq \label{phi}
\phi(x^\alpha; \tilde x^\alpha)= q \sum_{l=0}^\infty
\left(2l+1\right) P_l(\cos\gamma) g_l(\rho,\tilde \rho),
\eeq
where $\cos \gamma \equiv \cos \theta \cos \tilde \theta
+\sin \theta \sin \tilde \theta \cos(\varphi-\tilde \varphi)$
and $g_l(\rho,\tilde \rho)$ satisfies the equation
\beq \label{gl}
 g_l'' + \frac{{(r^2)}'}{r^2}g_l' - \left[ \frac{l(l+1)}{r^2} + \xi
R\right]g_l = -\frac{\delta(\rho, \tilde \rho)}{r^2}.
\eeq
In this expression and below a prime denotes a derivative with respect to $\rho$.
The homogeneous solutions to this equation will be denoted by $p_l(\rho)$
and $q_l(\rho)$. $p_{ l}(\rho)$ is chosen to be the solution which is
well behaved at $\rho =-\infty$ and divergent at $\rho \rightarrow +\infty$.
$q_l(\rho)$ is chosen to be the solution which is divergent at
$\rho \rightarrow -\infty$ and well behaved at $\rho =\infty$. Thus
\beq \label{pqeq}
 \left\{\frac{d}{d \rho^2} + \frac{{(r^2)}'}{r^2}\frac{d}{d \rho}
 - \left[ \frac{l(l+1)}{r^2} + \xi
R\right]\right\}\left\{p_{\, l}(\rho) \atop q_l(\rho) \right\} = 0,
\eeq
\bear
g_l(\rho,\tilde \rho)= C_l p_{\, l}(\rho_<) q_l(\rho_>)
= C_l && \left[\frac{}{}
\Theta (\tilde \rho-\rho) p_{\, l}(\rho) q_l(\tilde \rho)
\right. \nn && \left.
-\Theta (\rho-\tilde \rho) p_{\, l}(\tilde \rho) q_l(\rho)\right],
\ear
where $\Theta(x)$ is the Heaviside step function, i.e.,
$\Theta(x) = 1$ for $x > 0$
and $\Theta(x) = 0$ for $x < 0$,
$C_l$ is a normalization constant which could be absorbed into
the definition of $p_{\, l}$ and $q_l$. Normalization of $g_l$ is achieved
by integrating (\ref{gl}) once with respect to $\rho$ from
$\tilde \rho - \delta$ to $\tilde \rho + \delta$ and letting $\delta
\rightarrow 0$. This results in the Wronskian condition
\beq \label{WC}
C_l \left( p_{\, l} \frac{d q_l}{d \rho}- q_l \frac{d p_{\, l}}{d \rho} \right)
=-\frac{1}{r^2}.
\eeq

The WKB approximation for the radial modes $p_{\, l}$ and $q_l$ is obtained
by the change of variables \cite{AHS:1995}
\bear \label{pq}
p_{\, l}&=&\frac{1}{\sqrt{2r^2W}} \exp\left({\int^{\rho}} W d \rho \right),
\nn
q_l&=&\frac{1}{\sqrt{2r^2W}} \exp \left({-\int^{\rho}} W d \rho \right).
\ear
Substitution of these expressions into (\ref{WC}) shows that the Wronskian
condition is obeyed if
\beq
C_l=1.
\eeq
Substitution into the mode equation
(\ref{pqeq}) gives the following equation for $W$:
\bear \label{Weq}
W^2&=&{ \frac{l(l+1)+2 \xi }{r^2}}+\frac{{\left( W^2 \right)}''}{4 W^2}
-\frac{5 {{\left( W^2 \right)}'}^2}{16 W^4}
\nn &&
+\frac{{(r^2)}''}{2 r^2}-\frac{{{(r^2)}'}^2}{4 r^4}
+\xi\left( -2\frac{{(r^2)}''}{r^2}+\frac{{{(r^2)}'}^2}{2 r^4} \right).
\ear
This equation can be solved iteratively when the metric
function $r^2(\rho)$ is slowly varying, that is,
     \beq \label{lwkb}
     \varepsilon_{\mbox{\tiny \sl WKB}}=L_{\star} /L \ll 1,
     \eeq
where
      \beq \label{Lst}
      L_{\star}(\rho) =\frac{r(\rho)}{\sqrt{2\xi}},
      \eeq
and $L$ is a characteristic scale of variation of $r(\rho)$:
       \beq \label{Lm}
       \frac{1}{L(\rho)}= \max \left \{ \left| \frac{r'}{r}  \right|, \
        \left| \frac{r'}{r} \sqrt{\left|\xi \right|} \right|, \
       \left| \frac{r''}{r}  \right|^{1/2}, \
       \left|\frac{r''}{r} \left|\xi \right|  \right|^{1/2}, \
       \left| \frac{r'''}{r}  \right|^{1/3}, \
       \left| \frac{r'''}{r} \left|\xi \right|^{3/2}  \right|^{1/3}, \
       \dots  \right \} .
       \eeq
We shall call the region of spacetime where the metric function $r(\rho)$
is slowly varying the long throat.

The zeroth-order WKB solution of Eq. (\ref{Weq}) corresponds
to neglecting terms with derivatives in this equation
       \beq \label{Wsol}
       W^2=\Omega 
       \cdot
       \left(\frac{}{}1+O(\varepsilon_{\mbox{\tiny \sl WKB}}^2)  \right),
       \eeq
where
        \beq
        \Omega\left(\rho, l+1/2\right)=\frac{l(l+1)+2 \xi }{r^2}
        =\frac{1}{r(\rho)^2}\left[
        \left(l+\frac12\right)^2+\mu^2\right],
        \eeq
and
        \beq
       \mu^2=2\xi-\frac14.
        \eeq
Let us stress that $\Omega$ is the exact solution of equation (\ref{Weq}) in
a spacetime with metric
$ds^2=-d t^2+d\rho^2+r_0^2(d\theta^2+\sin^2\theta\, d\varphi^2)$,
where $r_0$ is constant.
Below it is assumed that
       \bear
       \mu^2 > 0.
       \ear
Substituting the solution (\ref{Wsol}) into (\ref{pq}) and (\ref{phi}),
and neglecting terms of the second order and higher with respect to
$\varepsilon_{\mbox{\tiny \sl WKB}}$ we can obtain the
following expression for the zeroth-order WKB approximation for
$\phi(x^\alpha; \tilde x^\alpha)$ under the assumptions $\theta=\tilde
\theta, \varphi=\tilde \varphi$ and $\tilde \rho=\rho+\delta \rho > \rho$
\beq
\phi(\rho, \theta, \varphi; \tilde \rho, \theta, \varphi)=
\frac{q}{r(\rho) r(\tilde \rho)} \sum_{l=0}^\infty \left(l+\frac12\right)
\frac{\exp \left({-\int\limits^{\rho+\delta \rho}_{\rho}}
\sqrt{\Omega\left(\rho', l+\dst \frac12\right)} d \rho' \right)}
{ \sqrt[4]{\Omega \left(\rho, l+ \dst \frac12\right) \Omega \left(\tilde \rho, l
+\dst \frac12\right)}}.
\eeq
The sum over $l$ can be evaluated by using the Plana sum method
(see, for example, \cite{Popov})
\bear \label{plana}
\phi(\rho, \theta, \varphi; \tilde \rho, \theta, \varphi)&=&
\frac{q}{r(\rho) r(\tilde \rho)} \lim_{\epsilon \rightarrow 0}
\left\{
\int \limits^{\infty}_{\epsilon}
\frac{\exp \left({-\int^{\rho+\delta \rho}_{\rho}}
\sqrt{\Omega(\rho', x)} d \rho' \right)}
{ \sqrt[4]{\Omega(\rho, x) \Omega(\tilde \rho, x)}} \, x d x
\right. \nn && \left.
+\int \limits^{\epsilon}_{\epsilon-i \infty}
\frac{ \exp \left({-\int^{\rho+\delta \rho}_{\rho}}
\sqrt{\Omega(\rho', z)} d \rho' \right)}
{ \sqrt[4]{\Omega(\rho, z) \Omega(\tilde \rho, z)}
\left( 1+e^{i 2 \pi z} \right)} \, z d z
\right. \nn && \left.
-\int \limits^{\epsilon+i \infty}_{\epsilon}
\frac{ \exp \left({-\int^{\rho+\delta \rho}_{\rho}}
\sqrt{\Omega(\rho', z)} d \rho' \right)}
{ \sqrt[4]{\Omega(\rho, z) \Omega(\tilde \rho, z)}
\left( 1+e^{-i 2 \pi z} \right)} \, z d z
\right\}.
\ear
The first integral in this expression can be rewritten as follows
\bear
&&\int \limits^{\infty}_{0}
\frac{\exp \left({-\int^{\rho+\delta \rho}_{\rho}}
\sqrt{\Omega(\rho', x)} d \rho' \right)}
{ \sqrt[4]{\Omega(\rho, x) \Omega(\tilde \rho, x)}} \, x d x
\nn
&=&\sqrt{r(\rho) r(\tilde \rho)}\int \limits^{\infty}_{0}
\frac{x \exp\left(-\sqrt{x^2+\mu^2}\int^{\rho+\delta \rho}_{\rho}
d \rho'/r(\rho')\right)} {\sqrt{x^2+\mu^2}} \, d x
\nn &&
=\sqrt{r(\rho) r(\tilde \rho)}\frac{\exp\left(-\mu\int^{\rho+\delta \rho}
_{\rho} d \rho'/r(\rho')\right)}
{\int^{\rho+\delta \rho}_{\rho} d \rho'/r(\rho')}
\ear
and expanded in powers of $\delta \rho$
\bear
&&\int \limits^{\infty}_{0}
\frac{\exp \left({-\int^{\rho+\delta \rho}_{\rho}}
\sqrt{\Omega(\rho', x)} d \rho' \right)}
{ \sqrt[4]{\Omega(\rho, x) \Omega(\tilde \rho, x)}} \, x d x
\nn
&=&\frac{r(\rho)^2}{\delta \rho} \left[ 1+\left(\frac{d r(\rho)}{ d \rho}
-\mu \right)\frac{\delta \rho}{r(\rho)}+O\left( \delta \rho^2 \right) \right].
\ear
The next two integrals in (\ref{plana}) do not diverge at $\delta \rho
\rightarrow 0$
\bear
&&\lim_{\epsilon \rightarrow 0}
\left\{
\int \limits^{\epsilon}_{\epsilon-i \infty}
\frac{ \exp \left({-\int^{\rho+\delta \rho}_{\rho}}
\sqrt{\Omega(\rho', z)} d \rho' \right)}
{ \sqrt[4]{\Omega(\rho, z) \Omega(\tilde \rho, z)}
\left( 1+e^{i 2 \pi z} \right)} \, z d z
\right. \nn && \left.
-\int \limits^{\epsilon+i \infty}_{\epsilon}
\frac{ \exp \left({-\int^{\rho+\delta \rho}_{\rho}}
\sqrt{\Omega(\rho', z)} d \rho' \right)}
{ \sqrt[4]{\Omega(\rho, z) \Omega(\tilde \rho, z)}
\left( 1+e^{-i 2 \pi z} \right)} \, z d z
\right\}
\nn &&
=r(\rho)\lim_{\epsilon \rightarrow 0}
\left\{
\int \limits^{i\epsilon+\infty}_{i\epsilon}
\frac{x d x}{ \sqrt{\mu^2-x^2}\left( 1+e^{ 2 \pi x} \right)}
\right. \nn && \left.
+ \int \limits^{-i\epsilon+ \infty}_{-i\epsilon}
\frac{x d x}{ \sqrt{\mu^2-x^2}\left( 1+e^{ 2 \pi x} \right)}
+O\left( \delta \rho \right)
\right\}
\nn &&
=2 r(\rho) \int^\mu_0 \frac{x d x}{\sqrt{\mu^2-x^2} \left( 1+e^{2 \pi x} \right)}
+O\left( \delta \rho \right).
\ear
Thus the zeroth-order WKB approximation of $\phi$ is
\bear
\phi(\rho, \theta, \varphi; \tilde \rho, \theta, \varphi)&=&
 \frac{q}{\delta \rho}
+\frac{q}{r(\rho)} \left( -\mu
+2 \int^\mu_0 \frac{x d x}{\sqrt{\mu^2-x^2} \left( 1+e^{2 \pi x} \right)}
\right)
\nn &&
+O\left(\frac{}{} \delta \rho \right).
\ear
The DeWitt-Schwinger counterterm $\phi_{\mbox{\tiny \sl DS}}(x; \tilde x)$
in the limit $\theta= \tilde \theta, \varphi=\tilde \varphi$
can be easily calculated using the metric (\ref{metric}):
\bear
&&2 \sigma = \delta \rho^2, \triangle = 1+O\left(\delta \rho^2\right),
\nn
&& \phi_{\mbox{\tiny \sl DS}}(\rho, \theta, \varphi; \tilde \rho, \theta, \varphi)
=q \frac{\triangle^{1/2}}{\sqrt{\, 2\sigma}}
=q\left(\frac{1}{\delta \rho}+O\left( \delta \rho \right) \right).
\ear
Thus $\phi_{\mbox{\tiny \sl ren}}(x)$ is
\bear \label{phiren}
\phi_{\mbox{\tiny \sl ren}}(x)&=&\lim_{ \delta \rho \rightarrow 0}
\left[ \phi(\rho, \theta, \varphi; \tilde \rho, \theta, \varphi)
- \phi_{\mbox{\tiny \sl DS}}(\rho, \theta, \varphi; \tilde \rho, \theta, \varphi)\right]
\nn
&=&\frac{q}{r(\rho)} \left( -\sqrt{2\xi-\frac14}
+2 \int\limits^{\sqrt{2\xi-1/4}}_0 \frac{x d x}
{\left( 1+e^{2 \pi x} \right)\sqrt{{2\xi-1/4}-x^2} }
\right)
\nn &&
\cdot \left(\frac{}{}1
+O(\varepsilon_{\mbox{\tiny \sl WKB}}^2)  \right),
\ear
and the single nonzero component of the self-force is
\bear \label{frho}
f_\rho(x)&=&-\frac{q}{2}\frac{\partial \phi_{\mbox{\tiny \sl ren}}}{\partial \rho}=
-\frac{q^2}{2r^2}\frac{d r}{d \rho} \left( \sqrt{2\xi-\frac14}
\right. \nn && \left.
-2 \int\limits^{\sqrt{2\xi-1/4}}_0 \frac{x d x}
{\left( 1+e^{2 \pi x} \right)\sqrt{{2\xi-1/4}-x^2} }
\right) \left(\frac{}{}1+O(\varepsilon_{\mbox{\tiny \sl WKB}}^2)  \right).
\ear
In the case $\xi =1/6$ we can numerically evaluate
\beq
F(\xi)= \sqrt{2\xi-\frac{\dst1}{\dst4}}
-2 \int\limits^{\sqrt{2\xi-1/4}}_0 \frac{\dst x d x}
{\dst \left( 1+e^{2 \pi x} \right)\sqrt{{2\xi-1/4}-x^2} }
\eeq
as follows $F(1/6) \simeq 0.1723\dots$.

\begin{figure}[ht]
\vbox{ \hfil \scalebox{0.4} {\includegraphics*{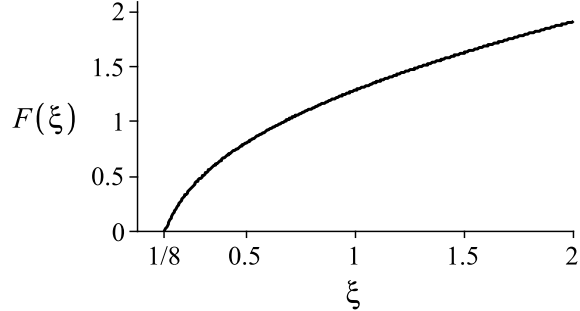}} \hfil }
\caption{The curve represents the function $ F(\xi) $. }
\end{figure}

Let us note that if one uses $r$ as the new radial coordinate
\beq
ds^2=-dt^2+\left(\frac{d \rho}{d r}\right)^2 d r^2+r^2 \left(d\theta^2+\sin^2\theta\, d\varphi^2\right),
\eeq
the expression (\ref{frho}) may be rewritten as follows
\beq \label{fr}
f_r= f_\rho \frac{d \rho}{d r}=-F(\xi)\frac{q^2}{2r^2}
\left(\frac{}{}1+O(\varepsilon_{\mbox{\tiny \sl WKB}}^2)  \right).
\eeq

\section{Specific examples}
First of all note that  $\varepsilon_{\mbox{\tiny \sl WKB}}=0$ in spacetime with metric
\beq \label{PH}
ds^2=-dt^2+d\rho^2+ {r_0}^2 \left(d\theta^2+\sin^2\theta\, d\varphi^2\right),
\eeq
where $r_0$ is constant and the expression (\ref{phiren}) is exact.
The self-force is zero in this case.

As a second example let us consider the spacetime with metric
\beq \label{m2}
ds^2=-dt^2+\frac{dr^2}{\dst \left( 1-\frac{r_g}{r} \right)^n}+
r^2 \left(d\theta^2+\sin^2\theta\, d\varphi^2\right).
\eeq
The case $n=2$ corresponds to the spacetime of a horn (semi-infinite throat) \cite{GHS:1991}.
The part of this spacetime with $r>r_g$ is globally static and geodesically
complete.
In the vicinity of $r=r_g$
\beq
\left|{\frac{d^m r}{r d \rho^m}}\right|^{1/m} \  \simeq  \ \frac{1}{r_g}\left(\frac{r-r_g}{r_g}\right)^{(n/2-1)+1/m},
\eeq
where $\rho$ is the radial proper distance
$\dst \left( dr/d\rho=\left( 1-{r_g}/{r} \right)^{n/2}\right)$.
Thus in the region $r-r_g \ll r_g$
\beq
\varepsilon_{\mbox{\tiny \sl WKB}} \simeq \left(\frac{r-r_g}{r_g}\right)^{n/2-1}.
\eeq
$\varepsilon_{\mbox{\tiny \sl WKB}} \ll 1$ in the case $n>2$
and one can call this region the long throat.

The expression (\ref{frho}) in the case $r-r_g \ll r_g$
and $n>2$ takes the form
\beq
f_\rho=-F(\xi)\frac{q^2}{2r(\rho)^2}\left( 1-\frac{r_g}{r(\rho)} \right)^{n/2}
\left(\frac{}{}1+O(\varepsilon_{\mbox{\tiny \sl WKB}}^2)  \right)
\eeq
or in the coordinates (\ref{m2})
\beq
f_r= f_\rho \frac{d \rho}{d r}=-F(\xi)\frac{q^2}{2r^2}
\left(\frac{}{}1+O(\varepsilon_{\mbox{\tiny \sl WKB}}^2)  \right).
\eeq

\section{Conclusions}

The considered approach gives the possibility to compute the exact
expression for the self-potential and the self-force in spacetime
(\ref{PH}). In the long throat (\ref{metric},\ref{lwkb}-\ref{Lm})
such approach permits to obtain the approximate expression for
the self-force (\ref{frho}, \ref{fr}). Let us note that the validity
of WKB approximation for all the modes (including $l=0$ mode)
of a massless scalar field is the consequence of the nonminimal
coupling ($\xi>1/8$) of a scalar field with the curvature of spacetime.
This implies also that the approximate solution (\ref{Wsol})
of the equation (\ref{Weq}) does not depend on the conditions at infinity
and in considered situation the effect of self-action is the local one.

\vspace{0.5cm} {\bf Acknowledgements:}

The author would like to thank N. Khusnutdinov for interesting and helpful
discussions.
This work was supported in part by grant 08-02-00325 from the Russian
Foundation for Basic Research.

\end{document}